\documentstyle[12pt,doublespace]{article}
\newcommand{\journal}[4]{{\em #1~}#2\,(19#3)\,#4;}

\newcommand{\hpa}{\journal {Helv. Phys. Acta}}

\newcommand{\ijmp}{\journal {Int. J. Mod. Phys.}}

\newcommand{\pr}{\journal {Phys. Rev.}}

\newcommand{\prl}{\journal {Phys. Rev. Lett.}}

\newcommand{\cmp}{\journal {Comm. Math. Phys.}}
\newcommand{\cqg}{\journal {Class. Quantum Grav.}}

\newcommand{\np}{\journal {Nucl. Phys.}}
\newcommand{\pl}{\journal {Phys. Lett.}}

\newcommand{\prep}{\journal {Phys. Reports}}

\newcommand{\nc}{\journal {Nuovo Cim.}}

\makeatletter
\@addtoreset{equation}{section}
\makeatother

\def\Lp{\displaystyle{\biggl(}}
\def\Rp{\displaystyle{\biggr)}}
\def\LP{\displaystyle{\Biggl(}}
\def\RP{\displaystyle{\Biggr)}}

\newcommand{\G}{\Gamma}
\newcommand{\D}{\Delta}
\renewcommand{\a}{\alpha}
\renewcommand{\b}{\beta}
\renewcommand{\d}{\delta}
\newcommand{\e}{\varepsilon}
\newcommand{\eb}{\bar\varepsilon}

\newcommand{\F}{\Phi}
\renewcommand{\P}{\Psi}
\newcommand{\Pb}{\bar\Psi}

\newcommand{\g}{\gamma}

\newcommand{\x}{\xi}
\renewcommand{\l}{\lambda}
\newcommand{\lb}{\bar\lambda}
\newcommand{\Lb}{\bar\Lambda}
\renewcommand{\L}{\Lambda}
\newcommand{\m}{\mu}
\newcommand{\n}{\nu}
\newcommand{\mn}{{\mu\nu}}
\newcommand{\mnrs}{\epsilon^{\mu\nu\rho\sigma}}
 \renewcommand{\O}{\Omega}
\newcommand{\p}{\psi}
\renewcommand{\pb}{\bar\psi}
\newcommand{\r}{\rho}
\newcommand{\s}{\sigma} \renewcommand{\S}{\Sigma}
\newcommand{\Sh}{\hat\Sigma}

\renewcommand{\AA}{{\cal A}}
\newcommand{\BB}{{\cal B}}
\newcommand{\BSh}{\hat{\cal B}_\Sigma}
\newcommand{\BS}{{\cal B}_\Sigma}

\newcommand{\FF}{{\cal F}}

\newcommand{\HH}{{\cal H}}

\newcommand{\NN}{{\cal N}}

\newcommand{\PP}{{\cal P}}
\newcommand{\QQ}{{\cal Q}}

\newcommand{\SS}{{\cal S}}

\newcommand{\identity}{\bf 1\hspace{-0.4em}1}
\newcommand{\complex}{{\kern .1em {\raise .47ex
\hbox {$\scriptscriptstyle |$}}
    \kern -.4em {\rm C}}}
\newcommand{\real}{{{\rm I} \kern -.19em {\rm R}}}
\newcommand{\rational}{{\kern .1em {\raise .47ex
\hbox{$\scripscriptstyle |$}}
    \kern -.35em {\rm Q}}}
\renewcommand{\natural}{{\vrule height 1.6ex width
.05em depth 0ex \kern -.35em {\rm N}}}

\newcommand{\cb}{{\bar c}}
\newcommand{\T}{(T^a)^A_B}
\newcommand{\half}{\frac 1 2}
\newcommand{\pa}{\partial}
\newcommand{\pad}[2]{{\frac{\partial #1}{\partial #2}}}
\newcommand{\fud}[2]  {{\displaystyle{\frac{\delta #1}{\delta #2}}}}

\newcommand{\ie}{{{\em i.e.}\ }}

\newcommand{\sla}{\raise.15ex\hbox{$/$}\kern -.57em}
\newcommand{\imply}{\quad\Rightarrow\quad}
\newcommand{\twiddle}{\lower.9ex\rlap{$\kern -.1em\scriptstyle\sim$}}

\newcommand{\gu}{(\eb^i\e_i)}
\newcommand{\gc}{(\eb^i\g_5\e_i)}
\newcommand{\gm}{(\eb^i\g^\m\e_i)}
\newcommand{\grii}{(\eb^i\g_R\e_i)}
\newcommand{\grij}{(\eb^i\g_R\e_j)}

\renewcommand{\=}{=&} 
\newcommand{\equ}[1]{(\ref{#1})}

\newcommand{\eq}{\begin{equation}}
\newcommand{\eqn}[1]{\label{#1}\end{equation}}
\newcommand{\eea}{\end{eqnarray}}
\newcommand{\eqap}{\begin{eqnarray}}
\newcommand{\eqanp}[1]{\label{#1}\end{eqnarray}}
\newcommand{\eqa}{\eq\ba{rl}}
\newcommand{\eqan}[1]{\ea\label{#1}\end{equation}}
\newcommand{\ba}{\begin{array}}
\newcommand{\ea}{\end{array}}
\newcommand{\eqac}{\begin{equation}\begin{array}{rcl}}
\newcommand{\eqacn}[1]{\end{array}\label{#1}\end{equation}}

\renewcommand{\pad}[2]{{\displaystyle{\frac{\partial #1}{\partial #2}}}}

\newcommand{\intx}{\int d^4 \! x \, }

\begin{document}
\def\ftoday{{\sl  \number\day \space\ifcase\month
\or Janvier\or F\'evrier\or Mars\or avril\or Mai
\or Juin\or Juillet\or Ao\^ut\or Septembre\or Octobre
\or Novembre \or D\'ecembre\fi
\space  \number\year}}



\titlepage
{\singlespace

\begin{center}

{\huge  Algebraic renormalization of N = 2 Super Yang--Mills theories
        coupled to matter }

\vspace{3cm}

{\Large Nicola Maggiore}\footnote{Supported in part
by the Swiss National Science Foundation.}

{\it D\'epartement de Physique Th\'eorique --
   Universit\'e de Gen\`eve\\24, quai E. Ansermet -- CH-1211 Gen\`eve
    4\\Switzerland}

\end{center}

\vspace{3.5cm}

\begin{center}
\bf ABSTRACT
\end{center}

{\it  We study the algebraic renormalization of $N=2$ Supersymmetric
 Yang--Mills theories coupled to matter. A regularization procedure
 preserving both the BRS invariance and the supersymmetry is not known
 yet, therefore it is necessary to adopt the algebraic method of
 renormalization, which does not rely on any regularization scheme. The
 whole analysis is reduced to the solution of cohomology problems
 arising from the generalized Slavnov operator which summarizes all the
 symmetries of the model. Besides to unphysical renormalizations of the
 quantum fields, we find that the only coupling constant of $N=2$ SYMs
 can get quantum corrections. Moreover we prove that all the
 symmetries defining the theory are algebraically anomaly--free.}

\vfill
\noindent
hep-th/9501057\\
UGVA-DPT-1994/12-876 \hfill December 1994
}
\newpage

\section{Introduction}

Since the early days of quantum field theory, physicists considered the
property of finiteness as one of the most appealing features a theory could
possess, according to the belief that the ultimate Theory of Nature
should be finite~\cite{js}. In the framework of quantum field theory, the
finiteness of a model results in the absence of quantum redefinitions
of the physical parameters -- and possibly of the quantum fields --,
meaning that the corresponding $\b$--functions -- together with the
anomalous dimensions related to the renormalizations of the quantum
fields --, vanish, which is also equivalent to say that the theory does
not exhibit ultraviolet divergences.

The supersymmetric field theories were the first to show a good
ultraviolet behaviour~\cite{ilzu}, and actually this represented the original
motivation for the interest arosen around them. In particular, the
most spectacular renormalization properties occurr in Supersymmetric
Yang--Mills
theories (SYMs) with extended $(N\geq 2)$ supersymmetry and calculations at
higher and higher loops~\cite{lps} strongly supported the conjecture that
the maximally extended one, namely $N=4$ SYM, was finite~\cite{sw1,conj}. Only
recently a formal proof of its finiteness to all orders of perturbations
theory has been given~\cite{whit1}.

As a matter of fact the $N=4$ case can be interpreted as a $N=2$ SYM with
matter in the adjoint representation of the gauge group and for this
reason the gauge field theories with extended $N=2$ supersymmetry are
considered as the most general ones~\cite{sw}. One important
property is that not every $N=2$ SYM is finite, but divergent quantum
corrections, if any, occurr only at one loop~\cite{sw1,west}. Moreover, the
physical relevance of $N=2$ SYMs has been stressed in a couple of
recent and authoritative papers~\cite{seiten}, which made them popular and
fashionable, thanks to some exact results
concerning very important topics like confinement and electric--magnetic
duality.

The study of the ultraviolet behaviour of a quantum field theory is
included in the renormalization program, which more generally concerns the
analysis of its divergence structure and the discussion of the
possible extension to the quantum level of the symmetries characterizing
it. When dealing with a supersymmetric theory, the absence of an
acceptable regularization procedure preserving both the BRS invariance
and the supersymmetry renders mandatory the adoption of the algebraic
method of renormalization, which indeed does not rely on any underlying
regularization scheme, and, moreover, leads to results valid to all
orders of perturbation theory.

The first algebraic study of the renormalizability of a supersymmetric
gauge field theory has been done for $N=1$ SYM in the
superfield formalism~\cite{ps1,ps2}, and recently the same results have
been recovered within the components description~\cite{whit2}.
Also the already cited proof of the finiteness of $N=4$ SYM~\cite{whit1}
has been given by using the algebraic method and adopting the
Wess--Zumino gauge, the superfield formalism presenting more difficulties
than advantages for the purpose of renormalizing field theories with
extended supersymmetry. The complete algebraic renormalization of the
general $N=2$ SYM is still lacking, the only attempt in that
direction being still uncompleted~\cite{bm}. The difficulties encountered
in~\cite{bm} originated from the infinite dimensional algebraic
structure, controlled at the price -- which resulted too high for the
renormalization of the model -- of introducing an infinite number of
external sources with increasing negative dimension.

Inspired by the algebraic proofs of the finiteness of the topological
quantum field theories~\cite{top}, which present analogous supersymmetric
algebraic structures, we gave in~\cite{1} a formulation of $N=2$ SYMs
alternative to that presented in~\cite{bm}, having the advantage of
being characterized by an algebra closed without making use neither of
equations of motion nor of auxiliary fields. The role of the latter is
indeed played by the external sources coupled to the nonlinear variations
of the quantum fields, and therefore necessarily present in the
theory~\cite{vari,gms}. The essence of the method followed in~\cite{1} was to
collect all the symmetries defining the theory into one generalized
Slavnov operator, so that the once complicated algebra reduced to a
simple nilpotency relation. In this paper we give the quantum extension
of the classical discussion made in~\cite{1}, which be briefly summarize
in Section 2. In Section 3 we perform the renormalization of the model,
which formally is that of an ordinary gauge field theory described by a
Slavnov identity~: first, we study the stability of the classical action
under radiative corrections and then we seek for possible anomalies. Some
conclusions are drawn in Section 4.

\section{The classical model}

In this section we briefly review the classical properties of the theory
which are necessary for what follows, referring to~\cite{1} for more
details. The fields of $N=2$ SYM are organized according to the vector
multiplet $(A^a_\m, \l_{\a i}, A^a, B^a)$, belonging to
the adjoint representation of a gauge group $G$, and the matter multiplet
$(A^{iA},A^\ast_{iA},\psi^\a_A,\bar\psi^A_\a)$, which is in an arbitrary
representation. In addition to these physical fields,
a ghost $c^a$, an antighost $\cb^a$ and a Lagrange multiplier $b^a$ are
introduced according to the usual gauge--fixing procedure.
The $N=2$ SYM is described by the complete gauge--fixed classical action
\eq
\S\equiv S_{inv} + S_{gf} + S_{ext}\ ,
\eqn{21}
where
\eq\ba{rl}
S_{inv}=& S_{SYM} + S_{matter} + S_{int} \\
=& \frac{1}{g^2} \intx \LP
    -\frac{1}{4}F^{a\m\n}F^a_{\m\n} +\frac{1}{2}(D^\m A)^a(D_\m A)^a
    +\frac{1}{2}(D^\m B)^a(D_\m B)^a
\\
&    +\half\lb^{ai}\g^\m(D_\m\l_i)^a
    +\half f^{abc}(\lb^{bi}\l^c_i)A^a
    -i\half f^{abc}(\lb^{bi}\g_5\l^c_i)B^a
    -\half f^{amn}A^mB^nf^{apq}A^pB^q \RP \\
&+ \intx\LP
(D^\m A^i)^A(D_\m A^\ast_i)_A
-\half\pb_A\g^\m(D_\m\p)^A\RP\\
&+\intx \LP
-(T^a)^A_B(\pb_A\l^a_i)A^{iB}
+(T^a)^A_B(\lb^{ai}\p^B)A^\ast_{iA}
+(T^a)^A_D(T^b)^D_BA^aA^bA^\ast_{iA}A^{iB}
\\
&+(T^a)^A_D(T^b)^D_BB^aB^bA^\ast_{iA}A^{iB}
-i\half (T^a)^A_B(\pb_A\g_5\p^B)B^a
-\half (T^a)^A_B(\pb_A\p^B)A^a\RP \ ,
\ea\eqn{22}
\eq
S_{gf} = \intx \LP b^a\partial A^a
- (\partial^\m\cb^a)(D_\m c)^a
   +(\pa^\m\cb^a)(\eb^i\g_\m\l^a_i) \RP \ ,
\eqn{23}
\eq\ba{rl}
S_{ext}= \intx\LP &
M^a(\QQ A^a) + N^a(\QQ B^a) + \O^{a\m}(\QQ A^a_\m)
+ \Lb^{ai}(\QQ\l^a_i) + L^a (\QQ c^a) \\
&
+ U^\ast_{iA}(\QQ A^{iA})
+ U^{iA}(\QQ A^\ast_{iA}) + \Pb_A(\QQ\p^A)
+ (\QQ\pb_A)\P^A
\\ &
- (\eb^j\L^a_i)(\Lb^{ai}\e_j) + \half (\eb^j\L^a_j)(\Lb^{ai}\e_i)
\\ &
- (\eb^i \e_i) (\Pb_A \P^A) + (\eb^i\g_5\e_i) (\Pb_A\g_5\P^A)
+ (\eb^i\g^\m\e_i) (\Pb_A\g_\m\P^A) \RP\ .
\ea\eqn{24}
Notice the absence of a bilinear term in the Lagrange multiplier $b$
in the gauge--fixing part $S_{gf}$ of the action, which
corresponds to choosing the Landau gauge.

In~\equ{21}, $S_{inv}+S_{gf}$ is left invariant by the operator $\QQ$ which
sums up BRS, supersymmetry and translations
by means of two ghost charged parameters $\e_i$
and $\x^\m$~:
\eq\ba{rl}
\QQ A^{a} \= f^{abc} c^b A^c + \eb^i\l^a_i + \x^\m\pa_\m A^{a} \\
\QQ  B^{a} \= f^{abc} c^b B^c + i\eb^i\g_5\l^a_i + \x^\m\pa_\m B^{a}
\\
\QQ A^{a}_\m \= -(D_\m c)^a + \eb^i\g_\m  \l^a_i + \x^\n\pa_\n A^{a}_\m
\\
\QQ \l^a_{\a i}   \= f^{abc} c^b \l_{\a i}^c +
           \frac{1}{2} F^a_{\m\n}(\s^{\m\n}\e_i)_\a
                  -(D_\m A)^a(\g^\m\e_i)_\a
\\ &                  +i(D_\m B)^a(\g^\m\g_5\e_i)_\a
                  +if^{abc}A^bB^c(\g_5\e_i)_\a + \x^\m\pa_\m \l^a_{\a i}
\\
\QQ A^{iA}    \= (T^a)^A_B c^a A^{iB} + \eb^i\psi^A + \x^\m\pa_\m A^{iA}
\\
\QQ \psi^A_\a \= (T^a)^A_B c^a \p^B_\a  -2(D_\m A^i)^A (\g^\m\e_i)_\a
                   +2(T^a)^A_B A^{iB} A^a \e_{\a i}
\\&
                -2i(T^a)^A_B A^{iB} B^a (\g_5\e_i)_\a + \x^\m\pa_\m \psi^A_\a
\\
\QQ c^a \= \half f^{abc}c^bc^c
- (\eb^i\g^\m\e_{i})A^a_\m - i(\eb^i\g_5\e_{i})B^a + (\eb^i\e_{i})A^a +
\x^\m\pa_\m c^a \\
\QQ \cb^a \= b^a + \x^\m\pa_\m \cb^a \\
\QQ b^a \= (\eb^i\g^\m\e_{i})\pa_\m\cb^a   + \x^\m\pa_\m b^a \\
\QQ \x^\m \= -(\eb^i\g^\m\e_{i}) \\
\QQ \e_i \= 0 \ .
\eqan{25}
The operator $\QQ$ is nilpotent provided that the spinor field equations
are satisfied
\eq
\QQ^2 = \mbox{equations of motion}\ ,
\eqn{26}
and this implies the presence in $S_{ext}$ of terms quadratic in the external
sources, in addition to the usual couplings
to the nonlinear $\QQ$--transformations
of the quantum fields. With such a source term in the total classical action
$\S$ it is possible to write the generalized Slavnov identity
\eqa
\SS(\S)\=\intx\LP
\fud{\S}{\O^{a\m}}
\fud{\S}{A^a_\m}
+
\fud{\S}{L^a}
\fud{\S}{c^a}
+
\fud{\S}{M^a}
\fud{\S}{A^a}
+
\fud{\S}{N^a}
\fud{\S}{B^a}
+
\fud{\S}{\Lb^{a i}}
\fud{\S}{\l^a_i}
\\&
+
\fud{\S}{U^\ast_{iA}}
\fud{\S}{A^{iA}}
+
\fud{\S}{U^{iA}}
\fud{\S}{A^\ast_{iA}}
+
\fud{\S}{\Pb_A}
\fud{\S}{\p^A}
+
\fud{\S}{\P^A}
\fud{\S}{\pb_A}
\\&
+
(b^a+\x^\m\pa_\m\cb^a)\fud{\S}{\cb^a}
+
( (\eb^i\g^\m\e_i)\pa_\m\cb^a+\x^\m\pa_\m b^a)\fud{\S}{b^a} \RP
- (\eb^i\g^\m\e_i)\pad{\S}{\x^\m}
\\
\= 0\ .
\eqan{27}
The corresponding linearized Slavnov operator
\eqa
\BS=\intx\LP &\!\!\!\!\!\!
\fud{\S}{\O^{a\m}}
\fud{}{A^a_\m}
+
\fud{\S}{A^a_\m}
\fud{}{\O^{a\m}}
+
\fud{\S}{L^a}
\fud{}{c^a}
+
\fud{\S}{c^a}
\fud{}{L^a}
+
\fud{\S}{M^a}
\fud{}{A^a}
+
\fud{\S}{A^a}
\fud{}{M^a}
\\&\!\!\!\!\!\!
+
\fud{\S}{N^a}
\fud{}{B^a}
+
\fud{\S}{B^a}
\fud{}{N^a}
+
\fud{\S}{\Lb^{a i}}
\fud{}{\l^a_i}
-
\fud{\S}{\l^a_i}
\fud{}{\Lb^{a i}}
+
\fud{\S}{U^\ast_{iA}}
\fud{}{A^{iA}}
+
\fud{\S}{A^{iA}}
\fud{}{U^\ast_{iA}}
\\&\!\!\!\!\!\!
+
\fud{\S}{U^{iA}}
\fud{}{A^\ast_{iA}}
+
\fud{\S}{A^\ast_{iA}}
\fud{}{U^{iA}}
+
\fud{\S}{\Pb_A}
\fud{}{\p^A}
-
\fud{\S}{\p^A}
\fud{}{\Pb_A}
+
\fud{\S}{\P^A}
\fud{}{\pb_A}
-
\fud{\S}{\pb_A}
\fud{}{\P^A}
\\&\!\!\!\!\!\!
+
(b^a+\x^\m\pa_\m\cb^a)\fud{}{\cb^a}
+
( (\eb^i\g^\m\e_i)\pa_\m\cb^a+\x^\m\pa_\m b^a)\fud{}{b^a} \RP
-(\eb^i\g^\m\e_i)\pad{}{\x^\m}
\eqan{28}
as a consequence of~\equ{27} is off--shell nilpotent
\eq
\BS\BS =0\ .
\eqn{29}

Finally, in addition to the Slavnov identity~\equ{27},
the classical theory is defined
by three further constraints~:
\begin{enumerate}
\item the gauge condition
\eq
\fud{\S}{b^a} = \pa A^a \ ,
\eqn{210}
whose commutator with the Slavnov identity~\equ{27} gives
the antighost equation
\eq
\bar\FF^a\S\equiv\fud{\S}{\cb^a} + \pa^\mu \fud{\S}{\O^{a\m}} -
\x^\m\pa_\m \fud{\S}{b^a} = 0\ ;
\eqn{211}
\item the $\x$--equation
\eq
\pad{\S}{\x^\m} = \D_\m\ ,
\eqn{212}
where
\eqa
\D_\m \equiv\intx \LP & \!\!\!
- M^a\pa_\mu A^a - N^a\pa_\mu B^a - \O^{a\n}\pa_\m A^a_\n
- (\Lb^{ai}\pa_\m\l^a_i) + L^a\pa_\m c^a
\\&\!\!\!
 - U^\ast_{iA}\pa_\m A^{iA}
- U^{iA}\pa_\m A^\ast_{iA} - (\Pb_A\pa_\m\p^A)
+(\pa_\m\pb_A\P^A)\RP\ ;
\eqan{213}
\item the ghost equation of the Landau gauge~\cite{bps,gms}
\eq
\FF^a\S=\D^a\ ,
\eqn{214}
where
\eq
\FF^a\equiv\intx\LP\fud{}{c^a}+f^{abc}\cb^b\fud{}{b^c}\RP
\eqn{215}
and
\eqa
\D^a \equiv \intx\LP&\!\!\!
f^{abc}\Lp M^bA^c + N^bB^c + \O^{b\m}A^c_\m + (\Lb^{bi}\l^c_i) - L^bc^c \Rp
\\&\!\!\!
-(T^a)^A_B\Lp U^\ast_{iA}A^{iB} - U^{iB}A^\ast_{iA}
+(\Pb_A\p^B) + (\pb_A\P^B)\Rp\RP\ .
\eqan{216}
\end{enumerate}

\section{Renormalization}

The proof of the renormalizability of the theory consists in showing that
it is possible to define a quantum vertex functional
\eq
\G=\S + O(\hbar)
\eqn{31}
which coincides at the lowest perturbative order with the classical
action~$\S$~\equ{21}, and which satisfies the generalized Slavnov identity
\eq
\SS(\G)=0\ .
\eqn{32}

The algebraic renormalization scheme
is performed according to two independent steps.
First we shall study the {\it stability}
of the classical action~$\S$ under radiative
corrections, checking that the most general invariant
counterterm can be reabsorbed through a redefinition of the fields
and  of the only coupling constant~$g^2$ of the theory. Then we shall discuss
the presence of {\it anomalies}, namely we shall investigate wether the
symmetries defining the theory can be implemented at the quantum
level.

This problem was addressed by Breitenlohner and Maison
in~\cite{bm}, but they encountered some difficulties originating from the
algebraic structure of $N=2$ SYMs. The supersymmetry algebra finds two
obstructions to the closure on the translations: equations of motion and
field dependent gauge transformations~\cite{1}. Following a standard
procedure, the authors of~\cite{bm} introduced auxiliary fields in order
to eliminate the equations of motion. Still, the
presence of the field dependent gauge transformations kept the algebra
infinite dimensional and therefore an infinite number
of external sources with increasing negative dimension were needed in order to
control the supersymmetric structure. This rendered the analysis of the
renormalization quite difficult and consequently a complete discussion
of the renormalization of $N=2$ SYM was never achieved.

The approach we are following here is different because the
classical Slavnov identity~\equ{27} has been obtained
in~\cite{1} by collecting
into an unique operator all the symmetries of the theory. Consequently, its
quantum extension corresponds to that of all the symmetries partecipating
in it, in particular the BRS transformations and the $N=2$ supersymmetry.
In other words, the absence of anomalies for the Slavnov identity~\equ{27}
implies that both the BRS symmetry and the N=2 supersymmetry are
anomaly--free as well~\cite{dix}.
Moreover, the study of the stability of the classical action
and of the anomalies technically reduces to
the analysis of cohomologies of the linearized Slavnov
operator~\cite{whit1,whit2,top,ms},
which is a far easier task than that encountered in~\cite{bm}.

\subsection{Stability of the classical action}

In order to find the most general invariant local counterterm, we perturb the
classical action
\eq
\S\longrightarrow\S+\eta\S_c\ ,
\eqn{33}
where $\eta$ is an infinitesimal parameter
and~$\S_c$ is the most general integrated local functional with canonical
dimension four and Faddeev--Popov ($\F\Pi$) charge
zero. We then require that the perturbed
action satisfies the symmetries defining the theory. At first order in~$\eta$
this corresponds to imposing  the following constraints on
the functional~$\S_c$~:
\begin{enumerate}
\item the gauge condition
      \eq
      \fud{}{b^a}(\S+\eta\S_c)=\pa A^a\imply\fud{\S_c}{b^a}=0\ ;
      \eqn{34}
\item the antighost equation
      \eq
      \bar\FF^a(\S+\eta\S_c) = 0\imply
      \fud{\S_c}{\cb^a} + \pa^\mu \fud{\S_c}{\O^{a\m}} =0\ ;
      \eqn{35}
\item the ghost equation
      \eq
      \FF^a(\S+\eta\S_c)=\D^a\imply\intx\fud{\S_c}{c^a}=0\ ,
      \eqn{36}
      where $\FF^a$ and $\D^a$ are given
      by \equ{215} and \equ{216} respectively;
\item the $\x$--equation
      \eq
      \pad{}{\x^\m}(\S+\eta\S_c) = \D_\m\imply\pad{\S_c}{\x^\m}=0\ ,
      \eqn{37}
      where $\D_\m$ is given by \equ{213};
\item the Slavnov identity~\equ{27}, which at first order in~$\eta$ implies
      the invariance of the perturbation~$\S_c$ under the action of the
      linearized Slavnov operator~\equ{28}
      \eq
      \SS(\S+\eta\S_c)=0\imply\BS\S_c=0\ .
      \eqn{38}
\end{enumerate}
The conditions \equ{34} and \equ{37} are satisfied by a functional which
does not depend neither on the Lagrange multiplier~$b^a$ nor on the
global ghost~$\xi^\m$. The antighost equation~\equ{35} implies that the
external source~$\O^{a\m}$ and the antighost~$\cb^a$ appear in~$\S_c$
only through the combination
\eq
\eta^{a\m}\equiv\pa^\m\cb^a+\O^{a\m}\ ,
\eqn{39}
while the effect of the ghost equation \equ{36} is that the perturbation
depends
on the ghost field~$c^a$ only if differentiated~$(\pa_\m c\equiv c_\m)$.
A functional satisfying all previous constraints depends on the fields
and parameters listed in Table 1 together with their quantum numbers.
\begin{center}
\begin{tabular}{|l|r|r|r|r|r|r|r|r|r|r|r|r|r|r|c|}\hline
 & $A_\mu^a$ & $\l_{\a i}^a$ & $A^a$ & $B^a$ & $A^{iA}$ & $\psi^\a_A$ &
                                       $c^a_\m$
 & $\eta^{a\m}$ & $\L_{\a i}$ & $M^a$ & $N^a$ & $U^{iA}$ & $\P^A$ & $L^a$ &
                                                                        $\e_i$
\\ \hline\hline
dim & $1$ & $3/2$  & $1$ & $1$ & $1$ & $3/2$ & $1$ & $3$ & $5/2$
    & $3$ & $3$   & $3$ & $5/2$ & $4$ & $-1/2$
\\ \hline
$\Phi\Pi$ & $0$ & $0$ & $0$ & $0$ & $0$ & $0$ & $1$ & $1$ & $1$
& $1$ & $1$ & $1$ & $1$ & $2$ & $1$
\\ \hline
\end{tabular}

\nopagebreak

\vspace{.1cm}

{\footnotesize {\bf Table 1. }
Dimensions and Faddeev--Popov
charges}
\end{center}

In the appendix we give the explicit form of the most general
perturbation~$\S_c$, by classifying it according to eigenstates of the
counting operator~$\NN_\e\equiv\eb^i\pad{}{\eb^i}$
\eq
\S_c=\S_c^{(0)}+\S_c^{(1)}+\S_c^{(2)}\ .
\eqn{310}
The counterterm $\S_c$ must satisfy the Slavnov condition~\equ{38},
which constitutes a cohomology problem, due to the nilpotency of
the linearized Slavnov operator~$\BB_\S$~\equ{28}. The general solution of
eq.~\equ{38} is
\eq
\S_c=\S_c^{(ph)} + \BB_\S\widehat\S_c\ .
\eqn{311}
A necessary condition for the renormalizability of the theory is that
the whole counterterm~$\S_c$ can be reabsorbed by a redefinition of the
quantum fields and of the coupling constant~$g^2$ of the classical action.
In particular,~$\widehat\S_c$ corresponds to unphysical field
renormalizations, called anomalous dimensions, while~$\S_c^{(ph)}$, which
cannot
be written as a $\BB_\S$--variation, entails a nonvanishing $\b$--function
of the coupling constant~$g^2$. Precisely, the physical renormalizations belong
to the cohomology sector with vanishing $\Phi\Pi$--charge of the linearized
Slavnov operator.

A long and straightforward calculation yields the following result for the
Slavnov condition~\equ{38}
\eq\ba{rl}
\S_c = & Z_{g^2}\LP S_{inv}
+ \intx \Lp
-(\eb^j\L^a_i)(\Lb^{ai}\e_j)
+\half (\eb^j\L^a_j)(\Lb^{ai}\e_i)
-(\eb^i\e_i)(\Pb_A\P^A)
\\&
+(\eb^i\g_5\e_i)(\Pb_A\g_5\P^A)
+(\eb^i\g^\m\e_i)(\Pb_A\g_\m\P^A)         \Rp \RP
\\&
+\BB_\S \intx \Lp c_1 \eta^{a\m}A^a_\m + c_2 M^aA^a
+ c_3 N^aB^a + c_4 \Lb^{ai}\l^a_i
\\&
+ c_5 (U^\ast_{iA}A^{ia} + U^{iA}A^\ast_{iA})
+ c_6  (\P_A\p^A + \pb_A\P^A) \Rp\ ,
\ea\eqn{312}
where $S_{inv}$ is the classical invariant action~\equ{22}, while $Z_{g^2}$
and $c_i$ are constants related to the renormalization of~$g^2$ and of
the fields respectively.

The celebrated finiteness property of the supersymmetric theories translates
either into the vanishing of the whole counterterm due to the algebraic
conditions on it -- as it happens for the topological models~\cite{top},
or into the lack of the physical part of it~$\S_c^{(ph)}$~\cite{ps3}.
Here on the contrary we find that $N=2$ SYMs exhibit a possible renormalization
of the coupling constant, besides to anomalous dimensions for the quantum
fields belonging to the vector and matter multiplet. Notice also that, as in
ordinary gauge field theories built in the Landau gauge~\cite{iz}, the ghost
field $c$ does not renormalize because of the ghost
equation~\equ{36}~\cite{bps,gms}.

The result \equ{312} is the best one can obtain with the algebraic
method of renormalization, according to which any claim on the
coefficients appearing in the
counterterm must be supported by a non anomalous symmetry of the classical
action. On the other hand, within the superspace background field
formulation of extended
supersymmetry, it has been possible to show that the $N=2$ SYMs, whose
$\b(g)$ function is vanishing at one loop, are finite to all orders
of perturbation theory~\cite{west}, and an analogous nonrenormalization
theorem has been proved for $N=1$ SYM~\cite{ps3} by exploiting the fact
that the $R$--current, the supersymmetry current and the energy momentum
tensor are components of one superfield. Within the $N=2$ SYMs,
a particular role is played by the maximally extended $N=4$ case,
which can be interpreted as a $N=2$ theory with matter in the adjoint
representation of the gauge group. In that case, indeed, it has been possible
to prove algebraically the perturbative finiteness, by showing the absence
of the superconformal anomaly~\cite{whit1}

\subsection{Anomalies}

The symmetries characterizing the theory are acceptable for any purpose,
including for instance the determination of the counterterm, only if they
survive the process of quantization. In our framework, this entails the
possibility of writing the quantum Slavnov identity~\equ{32}. The standard
algebraic procedure in order to prove that quantum implementation is
to assume that the Slavnov identity gets broken at the quantum level
by an insertion
\eq
\SS(\G) = \D\cdot\G\ .
\eqn{313}
A fundamental information on the breaking is provided by
the Quantum Action Principle~\cite{qap}, which states that at the lowest
nonvanishing order in~$\hbar$ the insertion~$\D\cdot\G$ is an integrated
local functional~$\D$ with dimension four and $\Phi\Pi$--charge one
\eq
\D\cdot\G = \D + O(\hbar\D)\ .
\eqn{314}
It is easy to show the validity to all orders of perturbation theory of the
gauge condition~\equ{210}, the antighost equation~\equ{211}, the
$\x$--equation~\equ{212} and the ghost equation~\equ{214}~\cite{psbk}
\eq\ba{rl}
\fud{\G}{b^a} = \pa A^a &\ \ \bar\FF^a\G = 0 \\
\pad{\G}{\x^\m}=\D_\m   &\ \ \FF^a\G = \D^a\ .
\ea\eqn{315}
The following algebraic relations hold
\eq
\BB_\g\SS(\g)=0
\eqn{316}
\eq\ba{rl}
\fud{}{b^a}\SS(\g) - \BB_\g (\fud{\g}{b^a} -\pa A^a) = & \bar\FF^a\g \\
\bar\FF^a\SS(\g) + \BB_\g\bar\FF^a\g = & 0 \\
\pad{}{\x^\m}\SS(\g) + \BB_\g (\pad{\g}{\x^\m} - \D_\m) = & \PP_\m\g \\
\FF^a\SS(\g) + \BB_\g (\FF^a\g-\D^a) = & \HH^a_{rig}\g
\ea\eqn{317}
where $\g$ is a generic functional, $\PP_\m$ and $\HH^a_{rig}$ are the
Ward operator for translations and rigid gauge invariance respectively.
Substituting in~\equ{316} and~\equ{317} the generic functional~$\g$
with the quantum vertex functional~$\G$ satisfying the relations~\equ{315}
and assuming as valid to all orders
\eq
\PP_\m\G = \HH^a_{rig}\G =0\ ,
\eqn{318}
the algebra \equ{317} yields the following constraints on the lowest order
breaking of the quantum Slavnov identity
\eq
\fud{\D}{b^a} = \bar\FF^a\D = \pad{\D}{\x^\m}= \FF^a\D = 0\ ,
\eqn{319}
which are satisfied by a functional depending only on the fields
and parameters listed in Table 1. In addition to the constraints~\equ{319},
the breaking~$\D$ is subjected to the Wess--Zumino consistency
condition~\cite{wz} arising from~\equ{316}
\eq
\BB_\S\D=0\ .
\eqn{320}
The equation \equ{320} is a cohomology problem like the Slavnov
condition~\equ{38} for the stability of the theory. The difference is that
this time the solution must belong to the space of local integrated functionals
with canonical dimension four and $\Phi\Pi$--charge {\it one} instead of
zero. The most general functional obeying the Wess--Zumino condition is
\eq
\D = \AA + \BB_\S \widehat\D\ ,
\eqn{321}
$\AA$ being a closed and not exact form
\eq
\AA \neq \BB_\S\widehat\AA\ .
\eqn{322}
If $\AA$ is present, or, equivalently, if the cohomology of the linearized
Slavnov operator~$\BB_\S$ in the space of the solutions of equation~\equ{320}
is not empty, the breaking~$\D$ cannot be reabsorbed by a fine tuning of
the fields and parameters of~$\G$. The functional~$\AA$ is an anomaly,
namely an obstruction to the validity  of the quantum Slavnov
identity~\equ{32}.
On the contrary, if there is no anomaly $(\AA=0)$,
the equations~\equ{320} and~\equ{321}
imply that the Slavnov identity holds good at a fixed order in~$\hbar$ and
hence, by induction, at every order.

The equation \equ{320} can be solved using the method of spectral
sequences~\cite{dix} or by looking directly for its general solution.
Either way, the calculation, although quite laborious,
{\it per se} does not present particular
difficulties. We solved the cohomology
problem~\equ{320} by writing the candidate for the anomaly as
the most general integrated local functional with dimension four,
$\Phi\Pi$--charge one and depending on the fields listed in Table 1.
The resulting functional~$\D$ is the sum of a huge number of terms,
which is convenient to gather according to their eigenvalues
of the counting operator
$\NN_\e\equiv\eb^i\pad{}{\eb^i}$
\eq
\D=\sum_{n=0}^3\D^{(n)}\ ,
\eqn{323}
with
\eq
[\NN_\e,\D^{(n)}] = n \D^{(n)}\ .
\eqn{324}
Notice that for power counting reasons it is not possible to write a local
integrated functional with the right quantum numbers and having~$n\geq 4$
\eq
\D^{(n)}=0\ \ \ \ \ \mbox{for}\ n\geq 4\ .
\eqn{326bis}
We must act on the functional $\D$ with the operator $\BB_\S$, which
accordingly writes
\eq
\BB_\S=\sum_{n=0}^2 s^{(n)}\ ,
\eqn{325}
with
\eq
[\NN_\e,s^{(n)}] = n s^{(n)}\ .
\eqn{326ter}
The explicit form of $\D^{(n)}$ and $s^{(n)}$ is given in the appendix.

The Wess--Zumino consistency condition \equ{320}, splitted into eigenstates
of the operator~$\NN_\e$, reads
\eq\ba{rl}
s^{(0)} \D^{(0)}= & 0 \\
s^{(0)} \D^{(1)} + s^{(1)} \D^{(0)} = & 0 \\
s^{(0)} \D^{(2)} + s^{(1)} \D^{(1)} + s^{(2)} \D^{(0)} = & 0 \\
s^{(0)} \D^{(3)} + s^{(1)} \D^{(2)} + s^{(2)} \D^{(1)} = & 0 \\
s^{(1)} \D^{(3)} + s^{(2)} \D^{(2)} = & 0 \\
s^{(2)} \D^{(3)}= & 0 \\
\ea\eqn{326}
It is both an easy and  long algebraic exercise to solve the
equations~\equ{326} and to verify that finally the solutions~$\D^{(n)}$
are such that the whole breaking~$\D$ is a $\BB_\S$--variation
\eq
\D = \BB_\S \widehat\D\ .
\eqn{327}
Notice that the result \equ{327} states that for $N=2$ SYMs
there are no anomalies already at algebraic level, contrarily to what happens
for ordinary YM theories and for $N=1$ SYMs, where the Adler--Bardeen
anomaly~\cite{ab}, or its supersymmetric extension~\cite{ps1,absup},
obeys the consistency condition~\equ{320}, and whose absence is guaranteed by
the vanishing of its coefficient~\cite{nonren}.

Therefore the Slavnov identity~\equ{27} can be implemented to all orders
of perturbation theory. This, together with the form~\equ{312}
of the counterterm, implies the renormalizability of $N=2$ SYMs.

\section{Conclusions}

We proved the renormalizability of $N=2$ SYMs in a purely algebraic way,
\ie without assuming the existence of any regularization scheme. The most
general counterterm~\equ{312} compatible with all the symmetries of the
theory can be reabsorbed by renormalizations of the only coupling
constant~$g^2$ and of the quantum fields belonging to the vector and
matter multiplet, the ghost field not renormalizing as a
consequence of the Landau gauge choice. This algebraic result reflects
the fact that in general $N=2$ SYMs do get divergent quantum corrections.
It is known, on the other hand~\cite{sw1,west}, that only those theories
verifying the condition~\equ{41} are finite~:
\eq
\sum_\s m_\s T(R_\s) = C_2(G)\ ,
\eqn{41}
where $m_\s$ is the number of $N=2$ matter multiplets in the
representation~$R_\s$ and the Casimir and Dynkin indices~$C_2,\ T$ are
defined as usual by
\eq\ba{rl}
C_2(G)\d^{ab} = & f^{amn}f^{bmn} \\
T(R_\s)\d^{ab} = & \mbox{Tr}      (T^a T^b)\ .
\ea\eqn{42}
The selection rule~\equ{41} has been obtained following two arguments.
The first of them~\cite{sw1} makes use of a particular superfield
formulation of~$N=2$ theories regulated by introducing higher
derivatives, which do not regulate the one loop contribution to the
quantum vertex functional~$\G$. The second argument~\cite{west} works in the
superspace background field formalism of~$N$--extended supersymmetric
theories, where no contribution to~$\G$ above one loop is possible.
The nonrenormalization
condition~\equ{41} corresponds to the vanishing at one loop, and hence at all
orders, of the $\b$--function of the coupling constant, and it has not
been reproduced yet in the general framework of an algebraic analysis,
relying only on the principles of locality
and power counting. For this purpose, our work may be the starting point to
extend to $N=2$ SYMs the one--loop {\it criteria} given in~\cite{ps3}
for the finiteness of $N=1$ supersymmetric gauge theories.
Notice that the $N=4$ theory, interpreted as a
$N=2$ with matter in the adjoint representation, trivially satisfies the
condition~\equ{41}, since $m_\s=1\ ,\ T(R_\s)=C_2(G)$.

The second part of the renormalization of $N=2$ SYMs consisted in the
verification that none of the symmetries forming the supersymmetry
algebra are anomalous and therefore hold good also at the quantum level.
This result has been achieved by exploiting the formulation given in~\cite{1},
according to which the whole analysis is reduced to the solution of the
cohomology problem~\equ{320} of the generalized Slavnov operator~\equ{28}. In
particular no $N=2$ supersymmetric extension of the Adler--Bardeen gauge
anomaly does exist.

{\bf Acknowledgments}
A.~Blasi, O.~Piguet and M.~Porrati are gratefully acknowledged for
stimulating discussions during the preparation of this work.

\appendix
{\singlespace
\section{Appendix}
\subsection{Linearized Slavnov operator}

In the space of functionals depending on the fields listed in Table 1,
the linearized Slavnov operator is modified as follows~\cite{1}
\eqap
\BSh=\intx\LP &&\!\!\!\!\!
\fud{\Sh}{\eta^{a\m}}
\fud{}{A^a_\m}
+
\fud{\Sh}{A^a_\m}
\fud{}{\O^{a\m}}
+
\fud{\Sh}{L^a}
\fud{}{c^a}
+
\fud{\Sh}{c^a}
\fud{}{L^a}
+
\fud{\Sh}{M^a}
\fud{}{A^a}
+
\fud{\Sh}{A^a}
\fud{}{M^a}
\\&&\!\!\!\!\!
+
\fud{\Sh}{N^a}
\fud{}{B^a}
+
\fud{\Sh}{B^a}
\fud{}{N^a}
+
\fud{\Sh}{\Lb^{a i}}
\fud{}{\l^a_i}
-
\fud{\Sh}{\l^a_i}
\fud{}{\Lb^{a i}}
+
\fud{\Sh}{U^\ast_{iA}}
\fud{}{A^{iA}}
+
\fud{\Sh}{A^{iA}}
\fud{}{U^\ast_{iA}}
\nonumber\\&&\!\!\!\!\!
+
\fud{\Sh}{U^{iA}}
\fud{}{A^\ast_{iA}}
+
\fud{\Sh}{A^\ast_{iA}}
\fud{}{U^{iA}}
+
\fud{\Sh}{\Pb_A}
\fud{}{\p^A}
-
\fud{\Sh}{\p^A}
\fud{}{\Pb_A}
+
\fud{\Sh}{\P^A}
\fud{}{\pb_A}
-
\fud{\Sh}{\pb_A}
\fud{}{\P^A}\RP\nonumber
\eqanp{a1}

By filtrating with the counting operator $\NN_\e$, $\BSh$ decomposes
as~\equ{325}. Explicitely we have
\eqap
s^{(0)} A^a &=& f^{abc}c^b A^c \nonumber\\
s^{(0)} B^a &=& f^{abc}c^b B^c \nonumber\\
s^{(0)} A^a_\m &=& - (D_\m c)^a \nonumber\\
s^{(0)} \l^a_{\a i} &=& f^{abc}c^b \l^c_{\a i} \nonumber\\
s^{(0)} c^a &=& \half f^{abc}c^b c^c \nonumber\\
s^{(0)} A^{iA} &=& (T^a)^A_B  c^a A^{iB} \nonumber\\
s^{(0)} \p^A_\a &=& (T^a)^A_B  c^a \p^{B}_\a \nonumber\\
s^{(0)} L^a &=& f^{abc}c^bL^c + f^{abc}M^bA^c + f^{abc}N^bB^c - (D\eta)^a +
               f^{abc}(\Lb^{bi}\l^c_i) - (T^a)^A_BU^\ast_{iA}A^{iB}
\nonumber\\&&            -(T^a)^A_B(\Pb_A\p^B) + (T^a)^A_BU^{iB}A^\ast_{iA} -
               (T^a)^A_B(\pb_A\P^B) \\
s^{(0)} M^a &=& -\frac{1}{g^2}(D^2A)^a + \frac{1}{2g^2}f^{abc}(\lb^{bi}\l^c_i)
               -\frac{1}{g^2}f^{abc}B^bf^{cmn}A^mB^n
\nonumber\\&&  +[(T^a)^A_D(T^b)^D_B+(T^b)^A_D(T^a)^D_B]A^bA^\ast_{iA}A^{iB}
               - \half(T^a)^A_B(\pb_A\p^B) + f^{abc}c^bM^c \nonumber\\
s^{(0)} N^a &=& -\frac{1}{g^2}(D^2B)^a
-\frac{i}{2g^2}f^{abc}(\lb^{bi}\g_5\l^c_i)
               +\frac{1}{g^2}f^{abc}A^bf^{cmn}A^mB^n
\nonumber\\&&  +[(T^a)^A_D(T^b)^D_B+(T^b)^A_D(T^a)^D_B]B^bA^\ast_{iA}A^{iB}
               -i\half(T^a)^A_B(\pb_A\g_5\p^B) + f^{abc}c^bN^c \nonumber\\
s^{(0)} \eta^{a\m} &=& -\frac{1}{g^2}(D_\n F^{\m\n})^a +
                      \frac{1}{g^2}f^{abc}A^b(D^\m A)^c +
                      \frac{1}{g^2}f^{abc}B^b(D^\m B)^c -
                      \frac{1}{2g^2}f^{abc}(\lb^{bi}\g^\m\l^c_i)
\nonumber\\&&                   +(T^a)^A_BA^{iB}(D^\m A^\ast_i)_A -
                      (T^a)^A_BA^\ast_{iA}(D^\m A^i)^B -
                      \half(T^a)^A_B(\pb_A\g^\m\p^B) +
                      f^{abc}c^b\eta^{c\m}\nonumber\\
s^{(0)} \Lb^{a\a i} &=& - (D_\m\lb^i\g^\m)^{a\a i} - f^{abc}\lb^{b\a i}A^c
                       + if^{abc}(\lb^{bi}\g_5)^\a B^c - (T^a)^A_B\pb^\a_A
A^{iB}
\nonumber\\&&          +(T^a)^A_B(i\g_5C\p^B)^\a \epsilon^{ij}A^\ast_{jA}
                       + f^{abc}\Lb^{b\a i}c^c \nonumber\\
s^{(0)} U^{iA} &=& -(D^2A^i)^A +\T(\lb^{ai}\p^B)
                   +(T^a)^A_D(T^b)^D_B(A^aA^b+B^aB^b)A^{iB} -\T U^{iB}c^a
\nonumber\\
s^{(0)} \P^A_\a &=& \half(\g^\m D_\m\p)^A_\a +\T\l^a_{\a i}
                    +i\half\T(\g_5\p^B)B^a
\nonumber\\&&
                    +\half\T\p^B_\a A^a +\T c^a\p^B_\a \nonumber
\eqanp{a2}
\eqap
s^{(1)} A^a &=& \eb^i\l^a_i \nonumber\\
s^{(1)} B^a &=& i\eb^i\g_5\l^a_i \nonumber\\
s^{(1)} A^a_\m &=& \eb^i\g_\m\l^a_i \nonumber\\
s^{(1)} \l^a_{\a i} &=& \half F^a_{\m\n}(\s^{\m\n}\e_i)_\a -
                     (D_\m A)^a(\g^\m\e_i)_\a +
                     i(D_\m B)^a(\g^\m\g_5\e_i)_\a +
                     if^{abc}A^bB^c(\g_5\e_i)_\a \nonumber\\
s^{(1)} c^a &=& 0 \nonumber\\
s^{(1)} A^{iA} &=& \eb^i\p^A \nonumber\\
s^{(1)} \p^A_\a &=& -2(D_\m A^i)^A(\g^\m\e_i)_\a
                   +2(T^a)^A_BA^{iB}A^a\e_{\a i}
                   -2i(T^a)^A_BA^{iB}B^a(\g_5\e_i)_\a \nonumber\\
s^{(1)} L^a &=& 0 \\
s^{(1)} M^a &=& (D_\m\Lb^i)^a\g^\m\e_i -if^{abc}(\Lb^{bi}\g_5\e_i)B^c
               +2(T^a)^A_B(\Pb_A\e_i)A^{iB}
               +2(T^a)^A_BA^\ast_{iA}(\eb^i\P^B) \nonumber\\
s^{(1)} N^a &=& -i(D_\m\Lb^i)^a\g^\m\g_5\e_i +if^{abc}(\Lb^{bi}\g_5\e_i)A^c
               -2i(T^a)^A_B(\Pb_A\g_5\e_i)A^{iB}
\nonumber\\&&  -2i(T^a)^A_BA^\ast_{iA}(\eb^i\g_5\P^B) \nonumber\\
s^{(1)} \eta^{a\m} &=& (D_\n\Lb^i)^a\s^{\m\n}\e_i
                      +f^{abc}(\Lb^{bi}\g^\m\e_i)A^c
                      -if^{abc}(\Lb^{bi}\g^\m\g_5\e_i)B^c
\nonumber\\&&                    -2(T^a)^A_B(\Pb_A\g^\m\e_i)A^{iB}
                      -2(T^a)^A_BA^\ast_{iB}(\eb^i\g^\m\P^A) \nonumber\\
s^{(1)} \Lb^{a\a i} &=& M^a\eb^i +iN^a(\eb^i\g^5)^\a
                       +\eta^{a\m}(\eb^i\g^\m)^\a \nonumber\\
s^{(1)} U^{iA} &=& -2(\eb^i\g^\m D_\m\P)^A +2\T A^a(\eb^i\P^B) -2i\T
                   B^a(\eb^i\g_5\P^B)
\nonumber\\
s^{(1)} \P^A_\a &=&- U^{iA}\e_{\a i}   \nonumber
\eqanp{a3}
\eqap
s^{(2)} A^a &=& 0 \nonumber\\
s^{(2)} B^a &=& 0 \nonumber\\
s^{(2)} A^a_\m &=& 0 \nonumber\\
s^{(2)} \l^a_{\a i} &=& -2(\eb^j\L^a_i)\e_{\a j} +(\eb^j\L^a_j)\e_{\a i}
\nonumber\\
s^{(2)} c^a &=& -(\eb^i\g^\m\e_i)A^a_\m -i(\eb^i\g_5\e_i)B^a
               +(\eb^i\e_i)A^a \nonumber\\
s^{(2)} A^{iA} &=& 0 \nonumber\\
s^{(2)} \p^A_\a &=& -(\eb^i\e_i)\P^A_\a +(\eb^i\g_5\e_i)(\g_5\P^A)_\a
                   +(\eb^i\g^\m\e_i)(\g_\m\P^A)_\a \nonumber\\
s^{(2)} L^a &=& 0 \\
s^{(2)} M^a &=& (\eb^i\e_i)L^a \nonumber\\
s^{(2)} N^a &=& -i(\eb^i\g_5\e_i)L^a \nonumber\\
s^{(2)} \eta^{a\m} &=& -(\eb^i\g^\m\e_i)L^a \nonumber\\
s^{(2)} \Lb^{a\a i} &=& 0 \nonumber\\
s^{(2)} \P^A_\a &=& 0\nonumber
\eqanp{a4}

\subsection{Counterterm}

The most general candidate for the counterterm is
$\S_c=\S_c^{(0)}+\S_c^{(1)}+\S_c^{(2)}$, where
\eqap
\S_c^{(0)}&=&\intx\LP
a_1F^{\mn}F^{a\m\n} +a_2f^{abc}(\lb^{bi}\l^c_i)A^a
+a_3f^{abc}(\lb^{bi}\g_5\l^c_i)B^a +a_4\lb^{ai}\g^\m(D_\m\l_i)^a
\nonumber\\&&
+a_5(T^a)^A_B(\lb^{ai}\p^B)A^\ast_{iA} +a_6(T^a)^A_B(\pb_A\l^a_i)A^{iB}
+a_7(T^a)^A_BA^a(\pb_A\p^B)
\nonumber\\&&
+a_8(T^a)^A_BB^a(\pb_A\g_5\p^b)
+a_9\pb_A\g^\m(D_\m\p)^A +a_{10}(D^\m A)^a(D_\m A)^a
\\&&
+a_{11}(D^\m B)^a(D_\m B)^a +a_{12}f^{amn}A^mB^nf^{apq}A^pB^q
+a_{13}(D^\m A^i)^A(D_\m A^\ast_i)_A
\nonumber\\&&
+a_{14}(T^a)^A_D(T^b)^D_BA^aA^bA^\ast_{iA}A^{iB}
+a_{15}(T^a)^A_D(T^b)^D_BB^aB^bA^\ast_{iA}A^{iB} +a_{16}c^a_\m\eta^{a\m}
\nonumber\\&&
+a_{17}D^{abcd}A^aA^bA^cA^d +a_{18}D^{abcd}B^aB^bB^cB^d
\RP\nonumber
\eqanp{a5}
\eqap
\S_c^{(1)}&=&\intx\LP
b_1(\eb^i\l^a_i)M^a +b_2(\eb^i\g_5\l^a_i)N^a
+b_3(\eb^i\g_\m\l^a_i)\eta^{a\m}
+b_4(\Lb^{ai}\g^\m\e_i)(D_\m A)^a
\nonumber\\&&
+b_5(\Lb^{ai}\g^\m\g_5\e_i)(D_\m B)^a
+b_6(\Lb^{ai}\s^{\m\n}\e_i)F^a_{\m\n}
+b_7(\Lb^{ai}\e_i)\pa A^a
\\&&
+b_8f^{abc}(\Lb^{ai}\g_5\e_i)A^bB^c
+b_9(T^a)^A_B(\Lb^{ai}\e_i)A^{jB}A^\ast_{jA}
+b_{10}(T^a)^A_B(\Lb^{aj}\e_i)A^{jB}A^\ast_{iA} \RP\nonumber
\eqanp{a6}
\eqap
\S_c^{(2)}&=&\intx\LP
c_1(\eb^i\e_i)L^aA^a +c_2(\eb^i\g_5\e_i)L^aB^a
+c_3(\eb^i\g^\m\e_i)L^aA^a_\m
\\&&
+c_{4_R}(\eb^i\g_R\L^a_i)(\Lb^{aj}\g_R\e_j)
+c_{5_R}(\eb^i\g_R\L^a_j)(\Lb^{aj}\g_R\e_i)
+c_{6_R}(\eb^i\g_R\P^A)(\Pb_A\g_R\e_i) \RP\ ,\nonumber
\eqanp{a7}
where $\g_R\in\{\identity ,\g_5,\g_\m,\g_\m\g_5,\s_{\m\n}\}$ and
$D^{abcd}$ is the completely symmetric invariant tensor of rank four
\eq
D^{abcd}\equiv d^{nab}f^{ncd} +d^{nac}f^{ndb} +d^{nad}f^{nbc}
\eqn{a8}
\subsection{Anomaly}

The most general candidate for the anomaly is
$\D=\D^{(0)}+\D^{(1)}+\D^{(2)}+\D^{(3)}$, where
\eq
\D^{(0)}=
\a_0\intx\epsilon^{\m\n\r\s}c^a_\m\Lp d^{abc}(\pa_\n A^b_\r)A^c_\s
+ \frac{1}{12}D^{abcd}A^b_\n A^c_\r A^d_\s\Rp \nonumber\\
+s^{(0)}\S_c^{(0)}\nonumber
\eqn{a9}
\eqap
&&\D^{(1)} =\\
&&\intx\LP\a_1 (\Lb^{ai}\e_i)\pa^2c^a
+\a_2^{abc}(\Lb^{ai}\g^\m\e_i)c^b_\m A^c
+\a_3^{abc}(\Lb^{ai}\g^\m\g_5\e_i)c^b_\m B^c
\nonumber\\&&
+\a_4^{abc}(\Lb^{ai}\e_i)A^b_\m c^{c\m}
+\a_5^{abc}(\Lb^{ai}\s^{\m\n}\e_i)A^b_\m c^c_\n
+\a_6\T(\eb^i\g^\m\P^A)c^a_\m A^\ast_{iA}
\nonumber\\&&
+\a_7\T(\Pb_A\g^\m\e_i)c^a_\m A^{iB}
+\a_8(\eb^i\l^a_i)\pa^2A^a
+\a_9(\eb^i\g_5\l^a_i)\pa^2B^a
\nonumber\\&&
+\a_{10}(\eb^i\g^\m\l^a_i)\pa_\m\pa A^a
+\a_{11}(\eb^i\g^\m\l^a_i)\pa^2A^a_\m
+\a_{12}(\eb^i\p^A)\pa^2A^\ast_{iA}
\nonumber\\&&
+\a_{13}(\pb_A\e_i)\pa^2A^{iA}
+\a_{14}^{abc}(\eb^i\g^\m\l^a_i)(\pa_\m A^b)A^c
+\a_{15}^{abc}(\eb^i\g^\m\g_5\l^a_i)(\pa_\m A^b)B^c
\nonumber\\&&
+\a_{16}^{abc}(\eb^i\g^\m\g_5\l^a_i)A^b(\pa_\m B^c)
+\a_{17}^{abc}(\eb^i\l^a_i)(\pa A^b)A^c
+\a_{18}^{abc}(\eb^i\l^a_i)A^b_\m(\pa^\m A^c)
\nonumber\\&&
+\a_{19}^{abc}(\eb^i\s^{\m\n}\l^a_i)(\pa_\m A^b_\n)A^c
+\a_{20}^{abc}(\eb^i\s^{\m\n}\l^a_i)A^b_\n(\pa_\m A^c)
+\a_{21}\T(\eb^i\g^\m\p^A)(\pa_\m A^a)A^\ast_{iA}
\nonumber\\&&
+\a_{22}\T(\eb^i\g^\m\p^A)A^a(\pa_\m A^\ast_{iA})
+\a_{23}\T(\pb_A\g^\m\e_i)(\pa_\m A^a)A^{iB}
\nonumber\\&&
+\a_{24}\T(\pb_A\g^\m\e_i)A^a(\pa_\m A^{iB})
+\a_{25}^{abc}(\eb^i\g^\m\l^a_i)(\pa_\m B^b)B^c
\nonumber\\&&
+\a_{26}^{abc}(\eb^i\g_5\l^a_i)(\pa A^b)B^c
+\a_{27}^{abc}(\eb^i\g_5\l^a_i)A^b_\m(\pa^\m B^c)
\nonumber\\&&
+\a_{28}^{abc}(\eb^i\s^{\m\n}\g_5\l^a_i)(\pa_\m A^b_\n)B^c
+\a_{29}^{abc}(\eb^i\s^{\m\n}\g_5\l^a_i)A^b_\m(\pa_\n B^c)
\nonumber\\&&
+\a_{30}\T(\eb^i\g^\m\g_5\p^B)(\pa_\m B^a)A^\ast_{iA}
+\a_{31}\T(\eb^i\g^\m\g_5\p^B)B^a(\pa_\m A^\ast_{iA})
\nonumber\\&&
+\a_{32}\T(\pb_A\g^\m\g_5\e_i)(\pa_\m B^a)A^{iB}
+\a_{33}\T(\pb_A\g^\m\g_5\e_i)B^a(\pa_\m A^{iB})
\nonumber\\&&
+\a_{34}^{abc}(\eb^i\g^\m\l^a_i)(\pa_\m A^b_\n)A^c_\n
+\a_{35}^{abc}(\eb^a\g^\m\l^a_i)(\pa^\n A^b_\m)A^c_\n
\nonumber\\&&
+\a_{36}^{abc}(\eb^i\g^\m\l^a_i)A^b_\m(\pa A^c)
+\a_{37}^{abc}\mnrs(\eb^i\g_\m\l^a_i)(\pa_\n A^b_\r)A^c_\s
\nonumber\\&&
+\a_{38}\T(\eb^i\p^B)(\pa A^a)A^\ast_{iA}
+\a_{39}\T(\eb^i\p^B)A^a_\m(\pa^\m A^\ast_{iA})
\nonumber\\&&
+\a_{40}\T(\pb_A\e_i)(\pa A^a)A^{iB}
+\a_{41}\T(\pb_A\e_i)A^a_\m(\pa^\m A^{iB})
\nonumber\\&&
+\a_{42}\T(\eb^i\s^{\m\n}\p^B)(\pa_\m A^a_\n)A^\ast_{iA}
+\a_{43}\T(\eb^i\s^{\m\n}\p^B)A^a_\m(\pa_\n A^\ast_{iA})
\nonumber\\&&
+\a_{44}\T(\pb_A\s^{\m\n}\e_i)(\pa_\m A^a_\n)A^{iB}
+\a_{45}\T(\pb_A\s^{\m\n}\e_i)A^a_\m(\pa_\n A^{iB})
\nonumber\\&&
+\a_{46}\T(\eb^i\g^\m\l^a_i)(\pa_\m A^{iB})A^\ast_{iA}
+\a_{47}\T(\eb^i\g^\m\l^a_i)A^{iB}(\pa_\m A^\ast_{iB})
\nonumber\\&&
+\a_{48}^{abcd}(\eb^i\l^a_i)A^bA^cA^d
+\a_{49}^{abcd}(\eb^i\g_5\l^a_i)A^bA^cB^d
\nonumber\\&&
+\a_{50}^{abcd}(\eb^i\g^\m\l^a_i)A^b_\m A^cA^d
+\a_{51}(T^a)^A_D(T^b)^D_BA^aA^b(\eb^i\p^B)A^\ast_{iA}
\nonumber\\&&
+\a_{52}(T^a)^A_D(T^b)^D_B(\pb_A\e_i)A^{iB}A^aA^b
+\a_{53}^{abcd}(\eb^i\l^a_i)A^bB^cB^d
\nonumber\\&&
+\a_{54}^{abcd}(\eb^i\g^\m\g_5\l^a_i)A^b_\m A^cB^d
+\a_{55}(T^a)^A_D(T^b)^D_BA^aB^b(\eb^i\g_5\p^B)A^\ast_{iA}
\nonumber\\&&
+\a_{56}(T^a)^A_D(T^b)^D_B(\pb_A\g_5\e_i)A^{iB}A^aB^b
+\a_{57}^{abcd}(\eb^i\l^a_i)A^bA^{c\m}A^d_\m
\nonumber\\&&
+\a_{58}^{abcd}(\eb^i\s^{\m\n}\l^a_i)A^bA^c_\m A^d_\n
+\a_{59}(T^a)^A_D(T^b)^D_B(\eb^i\g^\m\p^B)A^\ast_{iA}A^a_\m A^b
\nonumber\\&&
+\a_{60}(T^a)^A_D(T^b)^D_B(\pb_A\g^\m\e_i)A^{iB}A^a_\m A^b
+\a_{61}(T^a)^A_D(T^b)^D_B(\eb^i\l^a_i)A^bA^{iB}A^\ast_{iA}
\nonumber\\&&
+\a_{62}^{abcd}(\eb^i\g_5\l^a_i)B^bB^cB^d
+\a_{63}^{abcd}({\eb^i\g^\m\l^a_i})A^b_\m B^cB^d
%
%
\nonumber\\&&
+\a_{64}(T^a)^A_D(T^b)^D_B(\eb^i\p^B)A^\ast_{iA}B^aB^b
+\a_{65}(T^a)^A_D(T^b)^D_B(\pb_A\e_i)A^{iB}B^aB^b
\nonumber\\&&
+\a_{66}^{abcd}(\eb^i\g_5\l^a_i)A^b_\m A^{c\m}B^d
+\a_{67}^{abcd}(\eb^i\s^{\m\n}\g_5\l^a_i)A^b_\m A^c_\n B^d
\nonumber\\&&
+\a_{68}(T^a)^A_D(T^b)^D_B(\eb^i\g^\m\g_5\p^B)A^\ast_{iA}A^a_\m B^b
+\a_{69}(T^a)^A_D(T^b)^D_B(\pb_A\g^\m\g_5\e_i)A^{iB}A^a_\m B^b
\nonumber\\&&
+\a_{70}(T^a)^A_D(T^b)^D_B(\eb^i\g_5\l^a_i)B^bA^\ast_{iA}A^{iB}
+\a_{71}^{abcd}(\eb^i\g^\m\l^a_i)A^b_\m A^{c\n}A^d_\n
\nonumber\\&&
+\a_{72}^{abcd}\mnrs(\eb^i\g_\m\l^a_i)A^b_\n A^c_\r A^d_\s
+\a_{73}(T^a)^A_D(T^b)^D_B(\eb^i\p^B)A^\ast_{iA}A^{a\m}A^b_\m
\nonumber\\&&
+\a_{74}(T^a)^A_D(T^b)^D_B(\pb_A\e_i)A^{iB}A^{a\m}A^b_\m
+\a_{75}(T^a)^A_D(T^b)^D_B(\eb^i\s^{\m\n}\p^B)A^\ast_{iA}A^a_\m A^b_\n
\nonumber\\&&
+\a_{76}(T^a)^A_D(T^b)^D_B(\pb_A\s^{\m\n}\e_i)A^{iB}A^a_\m A^b_\n
+\a_{77}(T^a)^A_D(T^b)^D_B(\eb^i\g^\m\l^a_i)A^b_\m A^\ast_{iA}A^{iB}
\nonumber\\&&
+\a_{78}A^aA^a(\eb^i\p^A)A^\ast_{iA}
+\a_{79}A^aA^a(\pb_A\e_i)A^{iA}
\nonumber\\&&
+\a_{80}A^aB^a(\eb^i\g_5\p^A)A^\ast_{iA}
+\a_{81}(\pb_A\g_5\e_i)A^{iA}A^aB^a
\nonumber\\&&
+\a_{82}(\eb^i\g^\m\p^A)A^\ast_{iA}A^a_\m A^a
+\a_{83}(\pb_a\g^\m\e_i)A^{iA}A^a_\m A^a
+\a_{84}(\eb^i\l^a_i)A^aA^{jA}A^\ast_{jA}
\nonumber\\&&
+\a_{85}(\eb^i\p^A)A^\ast_{iA}B^aB^a
+\a_{86}(\pb_A\e_i)A^{iA}B^aB^a
+\a_{87}(\eb^i\g^\m\g_5\p^A)A^\ast_{iA}A^a_\m B^a
\nonumber\\&&
+\a_{88}(\pb_A\g^\m\g_5\e_i)A^{iA}A^a_\m B^a
+\a_{89}(\eb^i\g_5\l^a_i)B^aA^\ast_{jA}A^{jA}
+\a_{90}(\eb^i\p^A)A^\ast_{iA}A^{a\m}A^a_\m
\nonumber\\&&
+\a_{91}(\pb_A\e_i)A^{iA}A^{a\m}A^a_\m
+\a_{92}(\eb^i\g^\m\l^a_i)A^a_\m A^\ast_{jA}A^{jA}
+\a_{93}(T^a)^A_D(T^b)^D_B(\eb^i\l^a_j)A^bA^\ast_{iA}A^{jB}
\nonumber\\&&
+\a_{94}(\eb^i\l^a_j)A^aA^\ast_{iA}A^{jA}
+\a_{95}(T^a)^A_D(T^b)^D_B(\lb^{ai}\e_j)A^bA^\ast_{iA}A^{jB}
+\a_{96}(\lb^{ai}\e_j)A^aA^\ast_{iA}A^{jA}
\nonumber\\&&
+\a_{97}(T^a)^A_D(T^b)^D_B(\eb^i\g_5\l^a_j)B^bA^\ast_{iA}A^{jB}
+\a_{98}(\eb^i\g_5\l^a_j)B^aA^\ast_{iA}A^{jB}
\nonumber\\&&
+\a_{99}(T^a)^A_D(T^b)^D_B(\lb^{ai}\g_5\e_j)B^bA^\ast_{iA}A^{jB}
+\a_{100}(\lb^{ai}\g_5\e_j)B^aA^\ast_{iA}A^{jA}
\nonumber\\&&
+\a_{101}(T^a)^A_D(T^b)^D_B(\eb^i\g^\m\l^a_j)A^b_\m A^\ast_{iA}A^{jB}
%
%
+\a_{102}(\eb^i\g^\m\l^a_j)A^a_\m A^\ast_{iA}A^{jA}
\nonumber\\&&
+\a_{103}(T^a)^A_D(T^b)^D_B(\lb^{aj}\g^\m\e_i)A^b_\m A^\ast_{jA}A^{iB}
+\a_{104}(\lb^{aj}\g^\m\e_i)A^b_\m A^\ast_{jA}A^{iA}
\nonumber\\&&
+\a_{105}T^{BC}_{AD}(\eb^i\p^A)A^\ast_{iB}A^\ast_{jC}A^{jD}
+\a_{106}T^{BC}_{AD}(\pb_B\e_i)A^{iA}A^\ast_{jC}A^{jD}
\nonumber\\&&
+\a_{107}(T^a)^A_B(\lb^{ai}\g^\m\e_j)(\pa_\m A^{jB})A^\ast_{iA}
+\a_{108}(T^a)^A_B(\lb^{ai}\g^\m\e_j)A^{jB}(\pa_\m A^\ast_{iA})
\nonumber\\&&
+\a_{109}(T^a)^A_B(\eb^i\g^\m\l^a_j)(\pa_\m A^{jB})A^\ast_{iA}
+\a_{110}(T^a)^A_B(\eb^i\g^\m\l^a_j)A^{jB}(\pa_\m A^\ast_{iA})
\nonumber\\&&
+\a_{111}f^{abc}(\eb^i\l^a_i)(\lb^{bj}\l^c_j)
+\a_{112}f^{abc}(\eb^i\g_5\l^a_i)(\lb^{bj}\g_5\l^c_j)
\nonumber\\&&
+\a_{113}d^{abc}(\eb^i\g^\m\g_5\l^a_i)(\lb^{bj}\g_\m\g_5\l^c_j)
+\a_{114}d^{abc}(\eb^i\s^{\m\n}\l^a_i)(\lb^{bj}\s_{\m\n}\l^c_j)
\nonumber\\&&
+\a_{115_R}(T^a)^A_B(\eb^i\g_R\l^a_i)(\pb_A\g_R\p^B) \RP\nonumber
\eqanp{a10}
\eqap
&&\D^{(2)}=
\\&&
\intx\LP
\b_1(\eb^i\g^\m\e_i)L^ac^a_\m
+\b_2(\eb^i\e_i)\eta^{a\m}(\pa_\m A^a)
+\b_3\gm M^a(\pa_\m A^a)
\nonumber\\&&
+\b_4\gc\eta^{a\m}(\pa_\m B^a)
+\b_5\gm N^a(\pa_\m B^a)
+\b_6\gm\eta^a_\m(\pa A^a)
\nonumber\\&&
+\b_7\gm\eta^{a\n}(\pa_\m A^a_\n)
+\b_8\gm\eta^{a\n}(\pa_\n A^a_\m)
+\b_9\mnrs(\eb^i\g_\m\e_i)\eta^a_\n(\pa_\r A^a_\s)
\nonumber\\&&
+\b_{10}\gu M^a(\pa A^a)
+\b_{11}\gc N^a(\pa A^a)
+\b_{12}\gm U^\ast_{jA}(\pa_\m A^{jA})
\nonumber\\&&
+\b_{13}\gm U^{jA}(\pa_\m A^\ast_{jA})
+\b_{14}d^{abc}\gm A^aA^b\eta^c_\m
+\b_{15}d^{abc}M^aA^bA^c
\nonumber\\&&
+\b_{16}d^{abc}\gc N^aA^bA^c
+\b_{17}^{abc}\gc M^aA^bB^c
+\b_{18}^{abc}\gu N^aA^bB^c
\nonumber\\&&
+\b_{19}^{abc}\gu\eta^{a\m}A^b_\m A^c
+\b_{20}^{abc}\gm M^aA^b_\m A^c
+\b_{21}\T\gu U^\ast_{jA}A^{jB}A^a
\nonumber\\&&
+\b_{22}\T\gu A^aU^{jB}A^\ast_{jA}
+\b_{23}d^{abc}\gm B^aB^b\eta^a_\m
+\b_{24}d^{abc}\gu M^aB^bB^c
\nonumber\\&&
+\b_{25}d^{abc}\gc N^aB^bB^c
+\b_{26}^{abc}\gc\eta^{a\m}A^b_\m B^c
+\b_{27}^{abc}\gm N^aA^b_\m B^c
\nonumber\\&&
+\b_{28}\T\gc B^aU^\ast_{jA}A^{jB}
+\b_{29}\T\gc U^{jB}A^\ast_{jA}B^a
\nonumber\\&&
+\b_{30}d^{abc}\gu M^aA^{b\m}A^c_\m
+\b_{31}d^{abc}\gc N^aA^{b\m}A^c_\m
+\b_{32}d^{abc}\gm\eta^a_\m A^{b\n}A^c_\n
\nonumber\\&&
+\b_{33}^{abc}\gm\eta^{a\n}A^b_\m A^c_\n
+\b_{34}f^{abc}\mnrs(\eb^i\g_\m\e_i)\eta^a_\n A^b_\r A^c_s
+\b_{35}\T\gm A^a_\m U^\ast_{jA}A^{jB}
\nonumber\\&&
+\b_{36}\T\gm A^a_\m U^{jB}A^\ast_{jA}
+\b_{37}\T\gu M^aA^\ast_{jA}A^{jB}
\nonumber\\&&
+\b_{38}\gc\T N^aA^\ast_{jA}A^{jB}
+\b_{39}\T\gm\eta^a_\m A^\ast_{jA}A^{jB}
\nonumber\\&&
+\b_{1_R}\grij(\Lb^{aj}\g_R\g^\m\pa_\m\l^a_i)
+\b_{2_R}\grii(\pb_A\g_R\g^\m\pa_\m\p^A)
\nonumber\\&&
+\b_{3_R}\grii(\pa_\m\pb_A\g^\m\g_R\P^A)
+\b_{4_R}^{abc}\grij(\Lb^{aj}\g_R\l^b_i)A^c
\nonumber\\&&
+\b_{5_R}\T\grii(\Pb_A\g_R\p^B)A^a
+\b_{6_R}\T\grii(\pb_A\g_R\P^B)A^a
\nonumber\\&&
+\b_{7_R}^{abc}\grij(\Lb^{aj}\g_R\g_5\l^b_i)B^a
+\b_{8_R}\T\grii(\Pb_A\g_R\g_5\p^B)B^a
\nonumber\\&&
+\b_{9_R}\T\grii(\pb_A\g_R\g_5\P^B)B^a
+\b_{10_R}^{abc}\grij(\Lb^{aj}\g_R\g^\m\l^b_i)A^c_\m
\nonumber\\&&
+\b_{11_R}\T\grii(\Pb_A\g_R\g^\m\p^B)A^a_\m
+\b_{12_R}\T\grii(\pb_A\g_R\g^\m\P^B)A^a_\m
\nonumber\\&&
+\b_{13_R}\T\grij(\Pb_A\g_R\l^a_i)A^{jB}
+\b_{14_R}\T\grij(\lb^{aj}\g_R\P^B)A^\ast_{iA}
\nonumber\\&&
+\b_{15_R}\T\grij(\pb_A\g_R\L^a_i)A^{jB}
+\b_{16_R}\T\grij(\Lb^{aj}\g_R\p^B)A^\ast_{iA} \nonumber\RP
\eqanp{a11}
\eqap
&&\D^{(3)}=
\\&&
\intx\LP
\d_1\gu(\eb^j\L^a_j)M^a
+\d_2\gc(\eb^j\g_5\L^a_j)M^a
+\d_3\gu(\eb^j\g_5\L^a_j)N^a
\nonumber\\&&
+\d_4\gc(\eb^j\L^a_j)N^a
+\d_5\gu(\eb^j\g^\m\L^a_j)\eta^a_\m
+\d_6\gc(\eb^j\g^\m\g_5\L^a_j)\eta^a_\m
\nonumber\\&&
+\d_7\gm(\eb^j\L^a_j)\eta^a_\m
+\d_8\gu(\eb^j\l^a_j)L^a
+\d_9\gc(\eb^j\g_5\l^a_j)L^a
\nonumber\\&&
+\d_{10}\gu(\eb^j\P_A)U^\ast_{jA}
+\d_{11}\gc(\eb^j\g_5\P^A)U^\ast_{jA}
+\d_{12}\gu(\Pb_A\e_j)U^{jA}
\nonumber\\&&
+\d_{13}\gc(\Pb_A\g_5\e_j)U^{jA} \nonumber\RP
\eqanp{a12}
}


\begin{thebibliography}{999999}
\bibitem{js}     J.Schwinger, \pr{82}{51}{914}
\bibitem{ilzu}   J.Iliopulos and B.Zumino, \np{B76}{74}{310}
                 W.Lang and J.Wess, \np{B81}{74}{249}
\bibitem{lps}    S.Ferrara and B.Zumino, \np{B79}{74}{413}
                 D.R.T.Jones, \pl{72B}{77}{199}
                 E.Poggio and H.Pendleton, \pl{72B}{77}{200}
                 O.Tarasov, A.Vladimirov and A.Yu, \pl{93B}{80}{429}
                 M.T.Grisaru, M.Rocek and W.Siegel, \prl{45}{80}{1063}
                 W.E.Caswell and D.Zanon, \np{B182}{81}{125}
\bibitem{sw1}    M.Sohnius and P.West, \pl{100B}{81}{45}
\bibitem{conj}   M.Grisaru and W.Siegel, \np{B201}{82}{292}
                 P.Howe, P.K.Townsend and K.Stelle, \np{B236}{84}{125}
\bibitem{whit1}  P.L.White, \cqg{9}{92}{413}
\bibitem{sw}     M.F. Sohnius, \prep {128}{85}{39}
                 P. West, {\it Introduction to supersymmetry and supergravity},
                 Word Scientific Publishing 1990;
\bibitem{west}   P.Howe, K.Stelle and P.West, \pl{124B}{83}{55}
\bibitem{seiten} N. Seiberg and E. Witten, {\it Nucl. Phys.} B426(1994)19
                 --ERRATUM: {\it ibidem} B430(1994)485--; \np{B431}{94}{484}
\bibitem{ps1}    O.Piguet and K.Sibold, \np{B247}{84}{484}
\bibitem{ps2}    O.Piguet and K.Sibold, \np{B248}{84}{336} \np{B249}{84}{396}
\bibitem{whit2}  P.L.White, \cqg{9}{92}{1663}
\bibitem{bm}     P. Breitenlohner and D. Maison,
                 {\it Renormalization of supersymmetric Yang--Mills theories},
                 in Cambridge 1985, Proceedings,
                 {\it Supersymmetry and its applications}, p.309;
                 and {\it N=2 Supersymmetric Yang--Mills theories
                 in the Wess--Zumino gauge}, in
                 {\it Renormalization of quantum field theories with
                 nonlinear field transformations}, proceedings, workshop,
                 Tegernsee, 1987, by P. Breitenlohner, (Ed.), D. Maison, (Ed.),
                 K. Sibold, (Ed.) Springer (1988) p.64;
\bibitem{top}    N. Maggiore and S.P. Sorella, \np{B377}{92}{236}
                 \ijmp {A8}{93}{929}
                 A. Blasi and N. Maggiore, \cqg {10}{93}{37}
                 C. Lucchesi and O. Piguet   \np {B381}{92}{281}
                 C. Lucchesi, O. Piguet and S.P. Sorella \np {B395}{93}{325}
\bibitem{1}      N. Maggiore, {\it Off--shell formulation of N=2 SYM theories
                 coupled to matter without auxiliary fields}, preprint
                 UGVA-DPT-94-12-870, hep-th/9412092;
\bibitem{vari}   C. Becchi, {\it The nonlinear sigma model}, in Tegernsee 1987
                 (see Ref.~\cite{bm});
                 P.S. Howe, U. Lindstron and P.L. White,\pl {246B}{90}{430}
                 G.Bonneau, {\it BRS renormalizations of some on--shell
                 closed algebras of symmetry transformations: I and II},
                 hep-th/9406030-31;
\bibitem{gms}    E. Guadagnini, N. Maggiore and S.P. Sorella,\pl {255B}{91}{65}
\bibitem{bps}    A. Blasi, O. Piguet and S.P. Sorella, \np{B356}{91}{154}
\bibitem{dix}    J. A. Dixon,  \cmp {140}{91}{169} \cqg {7}{90}{1511}
\bibitem{ms}     N. Maggiore and M. Schaden, \pr {D50}{94}{6616}
\bibitem{ps3}    O.Piguet and K.Sibold, \pl{177B}{86}{373} \ijmp{A1}{86}{913}
                 C.Lucchesi, O.Piguet and K.Sibold,
                 \pl{201B}{88}{241} \hpa{61}{88}{321}
\bibitem{iz}     C.Itzykson and J.B.Zuber, {\it Quantum Field Theory},
                 McGraw--Hill 1980;
\bibitem{qap}    J.H.Jowenstein, \pr{D4}{71}{2281} \cmp{24}{71}{1}
                 Y.M.P.Lam, \pr{D6}{72}{2145} \pr{D7}{73}{2943}
                 T.E.Clark and J.H.Lowenstein, \np{B113}{76}{109}
\bibitem{psbk}   O.Piguet and S.P.Sorella, {\it Algebraic
                 Renormalization}, Springer Verlag 1995, in press;
\bibitem{wz}     J.Wess and B.Zumino, \pl{49B}{74}{52}
\bibitem{ab}     S.L.Adler, \pr{117}{69}{2426}
                 J.S.Bell and R.Jackiw, \nc{60}{69}{47}
                 W.A.Bardeen, \pr{184}{69}{48}
\bibitem{absup}  L.Bonora, P.Pasti, M.Tonin, {\it Nucl.
                 Phys.}B261(1985)249 --ERRATUM: {\it ibidem}
                 B269(1986)745--; \np{B252}{85}{458} \pl{156B}{85}{341}
                 G.Girardi, R.Grimm and R.Stora, \pl{156B}{85}{203}
\bibitem{nonren} G.Costa, J.Julve, T.Marinucci and M.Tonin, \nc{38A}{77}{373}
                 C.M.Becchi, A.Blasi, B.Bandelloni and R.Collina,
                 \cmp{72}{80}{239}
                 O.Piguet and S.P.Sorella, \np{B381}{92}{373}
                 {\it Nucl. Phys.} B395(1993)(661).






\end{thebibliography}
\end{document}